\documentclass{PoS}
\usepackage{bm}

\title{Jet Production at Low and High $\bm{Q^2}$ and Determination of the Strong Coupling $\bm{\alpha_s}$ at H1}

\ShortTitle{Jet Production at Low and High $Q^2$ and Determination of the Strong Coupling $\alpha_s$ at H1}

\author{\speaker{Roman Kogler}%
        \thanks{On behalf of the H1 collaboration.}\\
       Max-Planck-Institute for Physics, Munich\\
       E-mail: \email{roman.kogler@desy.de}}

\abstract{
Two recent measurements of inclusive jet, 2-jet and 3-jet cross sections in deep-inelastic $e\!p$ scattering from the H1 collaboration are presented. The measurements are performed at low $5<Q^2<100$ GeV$^2$ and high $150 < Q^2 < 15000$ GeV$^2$ virtualities of the exchanged boson. The obtained cross sections are in good agreement with perturbative QCD calculations at NLO and are used to extract the strong coupling $\alpha_s(M_Z)$. The experimental precision is 0.6\% and 1.2\% for the high and low $Q^2$ regimes, respectively. The error on the obtained value of $\alpha_s(M_Z)$ is dominated by the theoretical uncertainties which are mostly due to missing higher orders. At low $Q^2$ the uncertainty due to terms beyond NLO amounts to 7.5\%. The determination of $\alpha_s$ at various scales shows the running of the strong coupling over a large range.
}

\FullConference{XVIII International Workshop on Deep-Inelastic Scattering and Related Subjects\\
		 April 19 -23, 2010\\
		 Convitto della Calza, Firenze, Italy}

\begin{document}

\section{Jet production in deep-inelastic scattering}

Jet production in $e\!p$ scattering provides stringent tests of perturbative QCD and an independent assessment of the gluon contribution to parton density functions (PDFs). In contrast to inclusive DIS, the Born contribution to jet cross sections is directly sensitive to the strong coupling $\alpha_s$, making them an important tool for its determination.    

In two recent analyses by the H1 collaboration jet production in the Breit frame is measured at low virtuality of the exchanged boson $5<Q^2<100$ GeV$^2$ \cite{JetsLowQ2} and at high virtuality $150 < Q^2 < 15000$ GeV$^2$ \cite{JetsHighQ2}. The inelasticity $y$ is in both cases restricted to the range $0.2<y<0.7$ and the jet transverse momenta $P_T$ are measured in the Breit frame. In order to suppress the Born contribution to inclusive neutral current $e\!p$ scattering, which is independent of $\alpha_s$, significant jet $P_T$ in the Breit frame is required. To ensure jets are well contained within the acceptance of the LAr calorimeter their pseudorapidities\footnote{The pseudorapidity $\eta$ is defined as $-\ln(\tan(\theta/2))$ where $\theta$ is the polar angle in the H1 coordinate system.} in the laboratory frame have to be within $-1.0 < \eta^{jet}_{lab}< 2.5$ with slightly stricter cuts for the high $Q^2$ measurement. Jets are found in the Breit frame using the longitudinally invariant $k_T$ algorithm with $R_0=1$ and the massless $P_T$ recombination scheme.

\begin{figure}
\centering
\includegraphics[trim = 0mm 0mm 0mm 1.2cm, clip, width=0.9\textwidth]{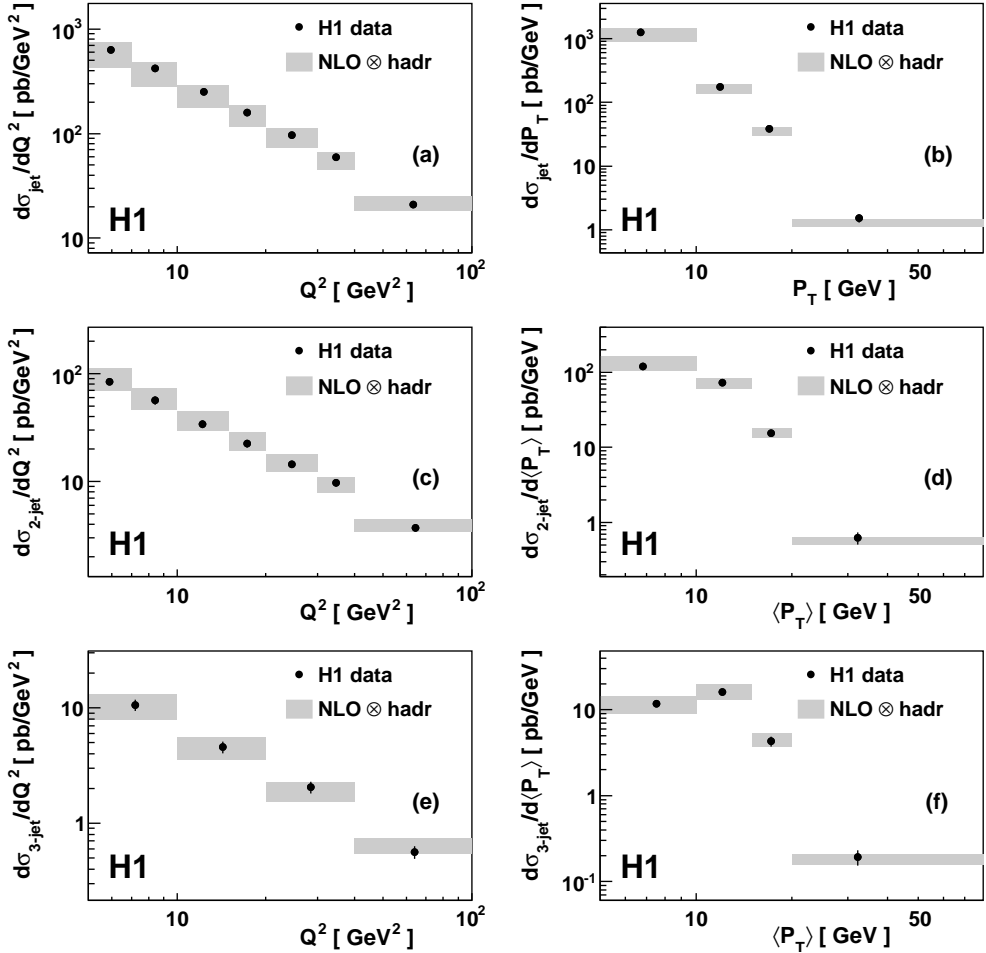}
\caption{Inclusive jet cross sections (a), (b), 2-jet cross sections (c), (d) and 3-jet cross sections (e), (f) as function of $Q^2$ (left) and $P_T$ or $\langle P_T \rangle$ (right). The experimental data are compared to NLO predictions corrected for hadronisation effects.}
\label{fig:jets_lowq2}
\end{figure}

In figure \ref{fig:jets_lowq2} inclusive jet, 2-jet and 3-jet cross sections as functions of $Q^2$ and $P_{T,obs}$ are shown for the low $Q^2$ measurement. $P_{T,obs}$ denotes the $P_T$ of the jet for inclusive jets and the average transverse momentum $\langle P_T \rangle$ of the two leading jets in 2- and 3-jet events. The data are compared to NLO calculations \cite{NLOJet} where the factorisation and renormalisation scales $\mu_f$ and $\mu_r$ were chosen to be $\sqrt{(Q^{2}+P_{T,obs}^2)/2}$. Different choices of $\mu_r$ involving only $Q^2$ or $P_{T,obs}$ are disfavoured by the data. The choice of the mixed scale for $\mu_r$ provides a good description of the data by the fixed-order calculation even at low $Q^2$ where $P_{T,obs}$ is the larger scale. Additionally it yields a smooth transition to the photoproduction regime with $Q^2\approx0$. The uncertainty on the NLO calculation due to missing higher orders is estimated by independently varying up and down $\mu_f$ and $\mu_r$ by a factor of 2. It is up to 3 times as large as the experimental uncertainty, which is dominated by the jet energy scale uncertainty and the uncertainty on the detector acceptance correction. 

For the high $Q^2$ measurement, jet cross sections are normalised to the inclusive DIS cross sections. In this case, $\mu_r$ has the same functional dependence as described above while the factorisation scale is set to $Q$ which is the natural choice for inclusive DIS. The normalised cross sections have smaller experimental uncertainties because of a partial cancellation of correlated uncertainties. The theoretical uncertainties are also reduced, owing to cancellations in the scale and PDF dependencies.

In leading order jet production the momentum fraction of the proton carried by the struck parton is $\xi=x_{Bj}(1+M_{12}^2/Q^2)$, with $x_{Bj}$ denoting the Bjorken scaling variable and $M_{12}$ the invariant mass of the two leading jets. At low $Q^2$ and $\xi$ boson-gluon fusion dominates jet production, making it directly sensitive to the gluon content of the proton. The 2-jet cross sections are measured as function of $\xi$ which can provide valuable input for PDF determinations.

\section{Determination of the strong coupling}

The strong coupling is obtained separately for the high and low $Q^2$ analyses from normalised and non-normalised cross sections, respectively. Fits are performed simultaneously to inclusive jet, 2-jet and 3-jet cross sections in bins of $Q^2$ and $P_{T,obs}$, yielding 54 and 62 data points for the high and low $Q^2$ measurements. For the latter measurement, data points with a $k$-factor above 2.5 are not considered in the fit where the $k$-factor is defined as the ratio of the NLO to LO cross sections. Large values of this ratio suggest a slow convergence of the perturbative series. The obtained values for $\alpha_s$ at the $Z^0$ mass are
\begin{equation}
\alpha_s(M_Z) = 0.1168 \pm 0.0007 (\mathrm{exp.})^{+0.0046}_{-0.0030} (\mathrm{th.}) \pm 0.0016 (\mathrm{pdf})
\label{eq:alphas_highq2}
\end{equation}
for the high $Q^2$ measurement and 
\begin{equation}
\alpha_s(M_Z) = 0.1160 \pm 0.0014 (\mathrm{exp.})^{+0.0093}_{-0.0077} (\mathrm{th.}) \pm 0.0016 (\mathrm{pdf})
\label{eq:alphas_lowq2}
\end{equation}
for the low $Q^2$ measurement. The smaller experimental uncertainty for the high $Q^2$ analysis (\ref{eq:alphas_highq2}) is due to the use of normalised cross sections. The theoretical uncertainty is the uncertainty on the hadronisation correction and the uncertainty arising from missing higher orders added in quadrature. The latter is obtained from a fit of $\alpha_s(M_Z)$ with $\mu_r$ and $\mu_f$ varied up and down by a factor of 2. At low $Q^2$, higher orders become more important and the theoretical uncertainty on $\alpha_s(M_Z)$ becomes twice as large as for the high $Q^2$ measurement. This yields a theoretical uncertainty which is about six times larger than the experimental uncertainty.

\begin{figure}
\centering
\includegraphics[width=0.4\textwidth]{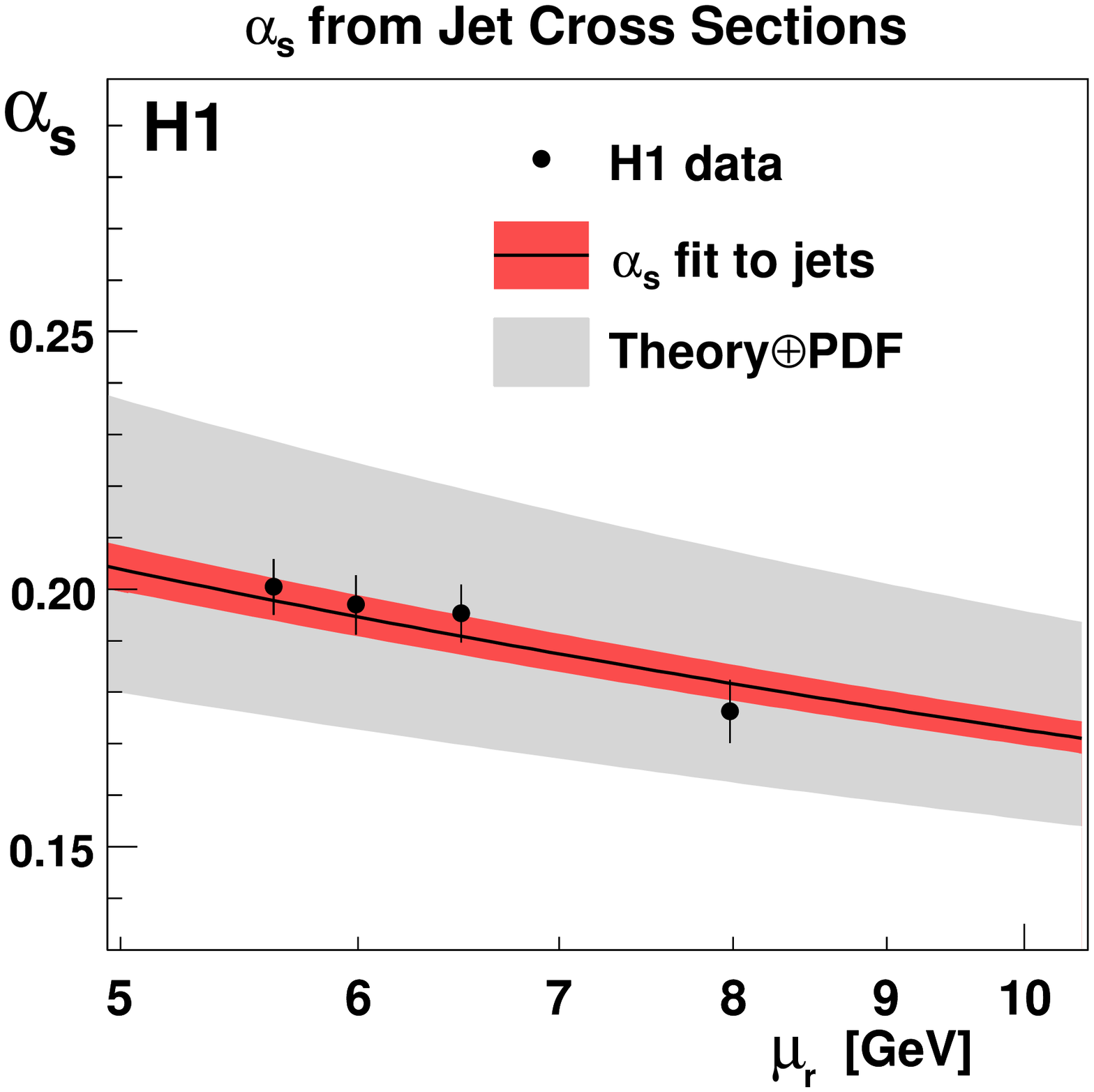} \put(-20,135){\small (a)}
\includegraphics[width=0.4\textwidth]{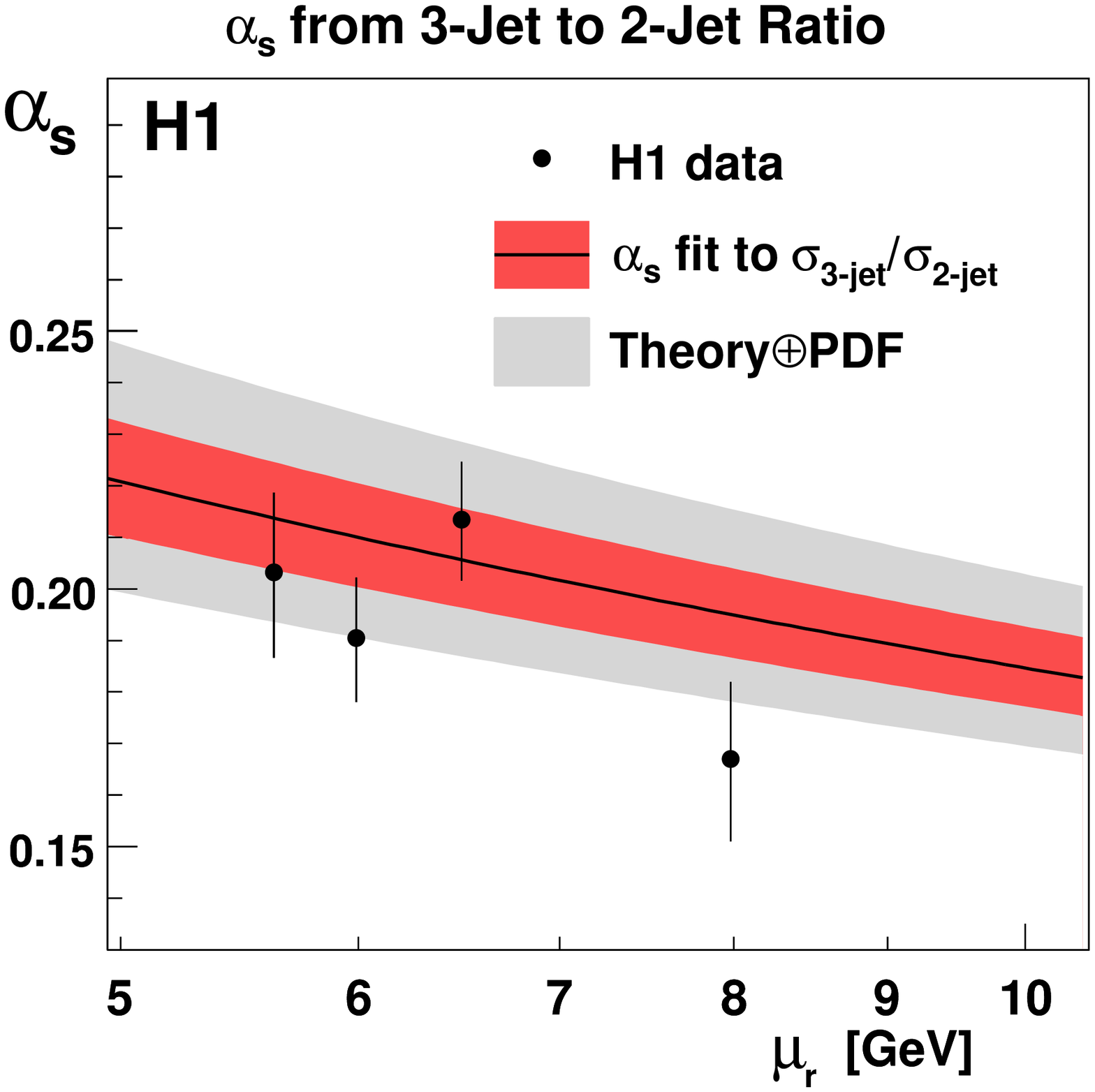} \put(-20,135){\small (b)}
\caption{Values of $\alpha_s(\mu_r)$ obtained from the low $Q^2$ measurement from a simultaneous fit to the inclusive, 2-jet and 3-jet cross sections (a) and from a fit to the 3-jet to 2-jet ratio (b). }
\label{fig:alphas_lowq2}
\end{figure}

Figure \ref{fig:alphas_lowq2}a shows values of $\alpha_s(\mu_r)$ obtained for each $Q^2$ bin from the low $Q^2$ analysis. The solid line indicates the two-loop solution of the renormalisation group equation (RGE) evolved from $\alpha_s(M_Z)$. The red and grey bands show the experimental and theoretical combined with the PDF uncertainties (\ref{eq:alphas_lowq2}), respectively. Using the 3-jet to 2-jet ratio $\sigma_{\rm{3-jet}}/\sigma_{\rm{2-jet}}$ for the determination of the strong coupling leads to a reduced theoretical uncertainty with the disadvantage of an increased statistical uncertainty (figure \ref{fig:alphas_lowq2}b). This could be improved by analysing the full HERA2 dataset.

For the extraction of $\alpha_s(M_Z)$ the CTEQ6.5M PDF is used and the uncertainty due to the choice of PDF is estimated using its error set. Alternatively to the Hessian method, the strong coupling has been extracted using the CTEQ6.6M PDF set which is obtained with different assumptions on $\alpha_s(M_Z)$. The result is compatible with the quoted values of $\alpha_s(M_Z)$ above, well within one standard deviation of the experimental uncertainty.

Obtained values of $\alpha_s(\mu_r)$ from both analyses are shown in figure \ref{fig:running_alphas}a together with the two-loop solution of the RGE using $\alpha_s(M_Z)$ from the high $Q^2$ analysis (\ref{eq:alphas_highq2}). Considering the large theoretical uncertainties in the low $Q^2$ regime the agreement is remarkable. The results from both measurements provide a test of the running of the strong coupling over a range from 6 to 70 GeV. A simultaneous fit to all data points from both analyses does not improve the precision on $\alpha_s(M_Z)$ due to the large theoretical uncertainties at low $Q^2$.

\section{Comparison with other measurements}

\begin{figure}
    \raisebox{-0.17cm}{\includegraphics[width=0.445\textwidth]{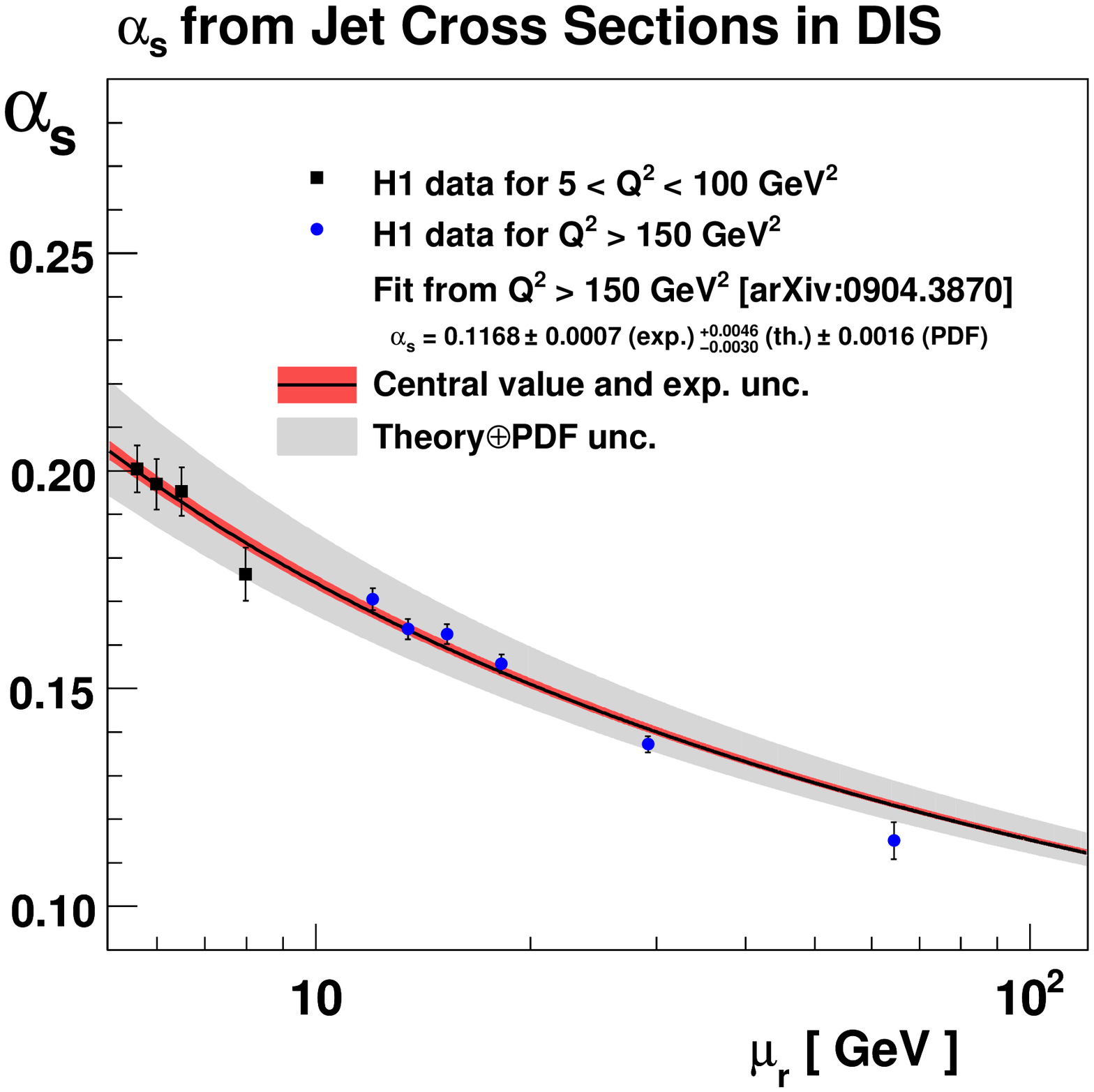}  \put(-20,150){\small (a)}}
    \includegraphics[width=0.56\textwidth]{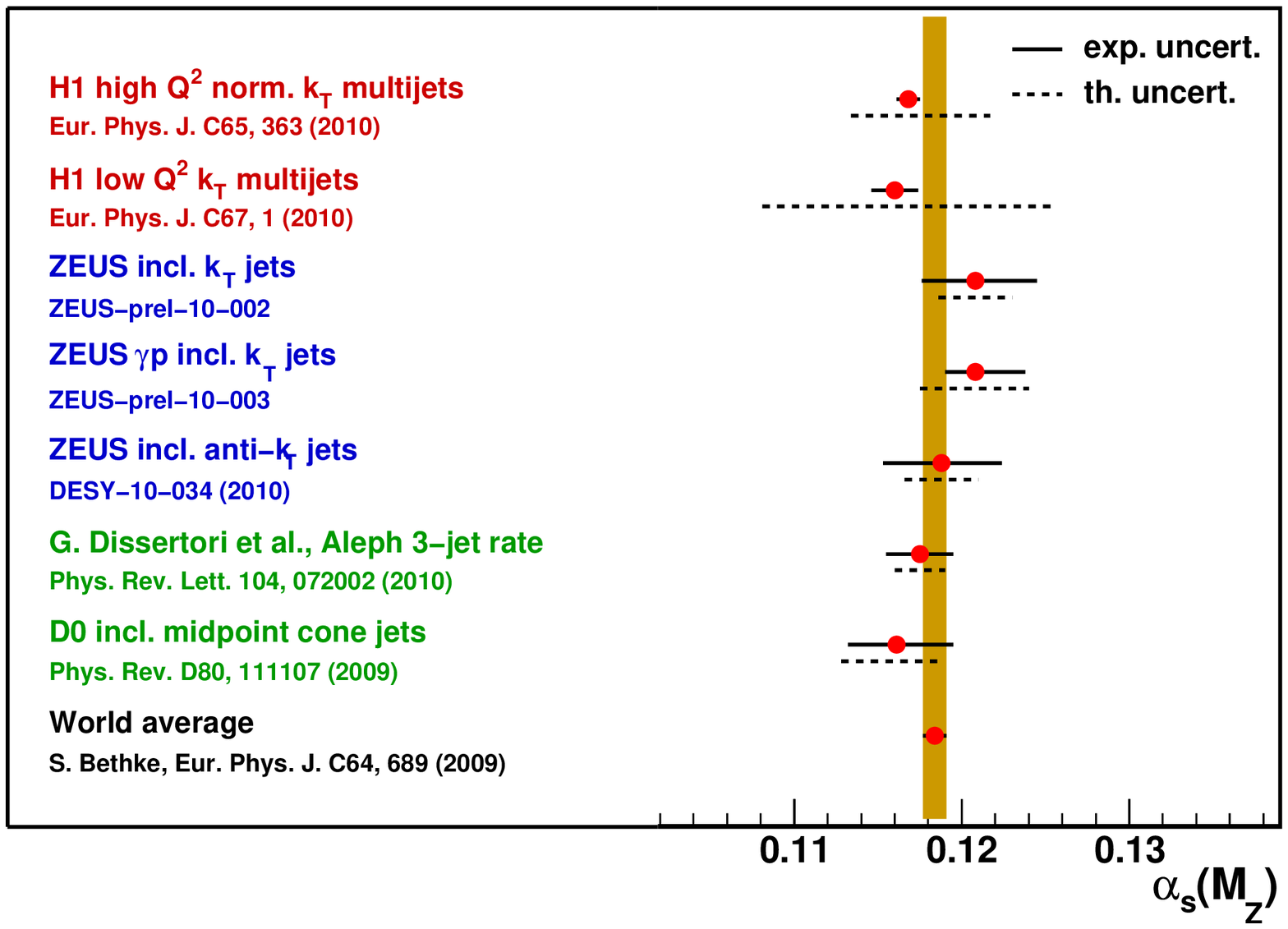}  \put(-28, 32){\small (b)}
  \caption{(a) Values of $\alpha_s(\mu_r)$ obtained from the measurement at low $Q^2$ (squares) and at high $Q^2$ (circles). (b) Recent $\alpha_s(M_Z)$ measurements obtained from jets in DIS, compared to the most precise measurements from jets in $e^+e^-$ annihilation and $p\bar{p}$ scattering. The world average is shown as coloured band.}
  \label{fig:running_alphas}
\end{figure}

A summary of recent measurements of $\alpha_s(M_Z)$ obtained from jet cross sections in deep-inelastic scattering is shown in figure \ref{fig:running_alphas}b. Also shown are the two most precise determinations of the strong coupling from jets in $e^+e^-$ annihilation \cite{Aleph} and in hadron-hadron collisions \cite{D0} together with the world average \cite{AlphasWorld}. The DIS jet measurements are competitive with those from other experiments and compatible with the world average. The experimental uncertainties from the H1 measurements are smaller than the ZEUS ones because of a simultaneous fit to inclusive jet, 2-jet and 3-jet cross sections, whereas the ZEUS extractions only use inclusive jet cross sections. At high $Q^2$ the use of normalised jet cross sections also significantly reduces the experimental uncertainty of the H1 measurement. The ZEUS method to estimate the uncertainty coming from missing terms beyond NLO yields a smaller theoretical uncertainty than the H1 method. Additionally, the ZEUS restriction of the fit range to $Q^2>500$ GeV$^2$ reduces the theoretical uncertainty with the drawback of increasing the experimental uncertainty. To make full use of the very precise jet data from deep-inelastic scattering NNLO calculations are essential, where the largest benefit would be at low $Q^2$.

\end{document}